\newcommand{\ourtitle}{My Model is Unfair, Do People Even Care?\\Visual Design Affects Trust and Perceived Bias in Machine Learning}
\newcommand{\ouronelinetitle}{My Model is Unfair, Do People Even Care? Visual Design Affects Trust and Perceived Bias in Machine Learning}

\documentclass[journal]{vgtc}                 
\ifpdf
  \pdfoutput=1\relax                   
  \usepackage{graphicx}                
  \DeclareGraphicsExtensions{.pdf,.png,.jpg,.jpeg} 
\else
  \usepackage{graphicx}                
  \DeclareGraphicsExtensions{.eps}     
\fi%

\graphicspath{{figures/}{pictures/}{images/}{./}} 

\usepackage{microtype}                 
\PassOptionsToPackage{warn}{textcomp}  
\usepackage{textcomp}                  
\usepackage{mathptmx}                  
\usepackage{times}
\usepackage{cite}                      
\usepackage{tabularx}                      
\usepackage{booktabs}                  

\usepackage{multirow}
\usepackage[table,xcdraw, svgnames]{xcolor}
\definecolor{salmon}{RGB}{232, 125, 114}
\definecolor{pink}{RGB}{232, 125, 114}
\definecolor{yuriy}{RGB}{34,139,34}

\newcommand{\pheading}[1]{\smallskip\noindent\textbf{#1}}
\usepackage{enumitem}
\onlineid{1073}

\vgtccategory{Theoretical and Empirical}

\title{\ourtitle}

\author{Aimen Gaba${}^\star\!\!$, Zhanna Kaufman${}^\star\!\!$, Jason Cheung, Marie Shvakel, Kyle Wm. Hall, Yuriy Brun, and Cindy Xiong Bearfield}

 \authorfooter{
  \item[${}^\star\!\!$] Aimen Gaba and Zhanna Kaufman contributed equally to this work as co-first authors.
  \item Aimen Gaba, Zhanna Kaufman, Jason Cheung, Marie Shvakel, Yuriy Brun, and Cindy Xiong Bearfield are with the University of Massachusetts, Amherst. E-mail: \{agaba, zhannakaufma, jasoncheung, mshvakel, brun, yaxiong\}@umass.edu.
  \item Kyle Hall is with Global Compliance, TD Bank. Email: kyle.hall@td.com.
 }

\shortauthortitle{Gaba and Kaufman, \MakeLowercase{\emph{et al.}}: \ouronelinetitle}

\usepackage{xargs}      
\usepackage{soul}       
\usepackage{color}      
\usepackage[table]{xcolor}      
\usepackage{xspace}     
\usepackage{xpunctuate} 
\usepackage{tcolorbox}
\usepackage{booktabs}
\usepackage{balance}

\definecolor{lightpink}{RGB}{237,157,202}
\definecolor{lightred}{RGB}{210,121,121}
\definecolor{lightorange}{RGB}{230,170,50}
\definecolor{lightgold}{RGB}{210,194,121}
\definecolor{lightgreen}{RGB}{121,210,121}
\definecolor{lightaqua}{RGB}{121,206,210}
\definecolor{lightblue}{RGB}{121,124,210}
\definecolor{lightpurple}{RGB}{153,102,255}
\definecolor{red}{RGB}{178,34,34}
\definecolor{gray}{RGB}{166,166,166}

\newcommand{\model}{model\xspace}

\newcommand{\eat}[1]{\relax}

\abstract{%
Machine learning technology has become ubiquitous, but, unfortunately, often exhibits bias. As a consequence, disparate stakeholders need to interact with and make informed decisions about using machine learning models in everyday systems. 
Visualization technology can support stakeholders in understanding and evaluating trade-offs between, for example, accuracy and fairness of models. 
This paper aims to empirically answer ``Can visualization design choices affect a stakeholder's perception of model bias, trust in a model, and willingness to adopt a model?''
Through a series of controlled, crowd-sourced experiments with more than 1,500 participants, we identify a set of strategies people follow in deciding which models to trust. 
Our results show that men and women prioritize fairness and performance differently and that visual design choices significantly affect that prioritization.
For example, women trust fairer models more often than men do, participants value fairness more when it is explained using text than as a bar chart, and being explicitly told a model is biased has a bigger impact than showing past biased performance. 
We test the generalizability of our results by comparing the effect of multiple textual and visual design choices and offer potential explanations of the cognitive mechanisms behind the difference in fairness perception and trust. 
Our research guides design considerations to support future work developing visualization systems for machine learning.
}

\keywords{machine learning, fairness, bias, trust, visual design, gender, human-subjects studies}

\begin{document}


\firstsection{Introduction}
\maketitle
\label{sec:Introduction}

Data-driven systems that use machine learning (ML) are ubiquitous in today's society, spanning high-impact domains such as healthcare~\cite{Komorowski18}, banking~\cite{qai22}, hiring~\cite{Roy20}, and the criminal justice system~\cite{Angwin16}.
Unfortunately, such systems can be unsafe and biased (e.g., racist or sexist), which erodes people's trust. 
For example, IBM Watson recommended potentially fatal cancer treatments~\cite{Ross18}, cancer diagnosis systems have exhibited lower detection rates for people of color~\cite{Yaya19}, software used by courts in setting bail have been found to have racial bias~\cite{Angwin16}, and facial recognition systems routinely discriminate against women and people of color~\cite{Buolamwini18}.
Such issues have led to legal bans of some types of ML systems~\cite{Singer19, Sheard22}. 
While extensive work focuses on reducing bias in ML algorithms~\cite{Thomas19, Giguere22iclr, 
bell2023possibility, Metevier19neurips}, 
such methods often result in compromises; for example, sacrificing system accuracy for fairness, or requiring more expensive data or computational resources, thereby necessitating human involvement and complex decision-making.

Visualization is one powerful strategy to inform users of such compromises~\cite {Cabrera19, Johnson23fairkit}, but visualization design choices can profoundly affect how people reason~\cite{xiong2022reasoning}, compare data values~\cite{gaba2022comparison, xiong2021perceptual}, infer about people~\cite{holder2022dispersion}, draw causal conclusions~\cite{xiong2019illusion}, trust the data~\cite{padilla2022multiple, xiong2019examining, mayr2019trust, elhamdadi2022trust}, and perceive fairness~\cite{berkel2021, wang2020}.
Therefore, practitioners who create visualizations to communicate ML model information must proceed cautiously with their design choices, as even without visualizations the way ML models are described can impact people's trust in those models~\cite{yin2019understanding} and how they perceive model fairness~\cite{berkel2021}.
As stakeholders with a variety of knowledge and experience use visualization to support reasoning about ML models~\cite{lai2021science}, the visualization community must study the effects of visualization on how people reason about ML models, including perceptions of model fairness and trustworthiness.

\begin{tcolorbox}
This paper addresses this underexplored space by empirically assessing how visual design, model performance and fairness, and user characteristics affect people's trust in ML models.
\end{tcolorbox}

We focused on the demographic parity aspect of fairness, i.e., the difference in positive outcomes across protected groups~\cite{Dwork12, Galhotra17fse}, and considered gender-based biases.\footnote{Gender is not binary. In this paper, we focus on bias against men and women. Future work should explore broader gender implications.} 
Inspired by trust research from behavioral economics~\cite{glaeser2000measuring, coleman1990foundations}, we performed a series of experiments to understand how visualization design choices can impact trust and perceived fairness in decision-making with ML model outputs.
We adopted a trust-game framework commonly used by economists to study trust~\cite{zurn2017trust}: we showed participants pairs of investment models (one fair and one biased), and they selected the model in which they would invest (i.e., entrust) their money. 
The commitment to invest serves as a proxy for trust, and the frequency of investing in the fair model encodes the relationship between perceived model performance and fairness. 
This trust game-based instrument allowed us to analyze how people's trust in models can be shaped by visualization design choices, the models' performance and fairness, and user characteristics. 

\begin{figure*}[t]
\rowcolors{1}{gray!25}{white}
{\renewcommand{\arraystretch}{1.6}
\begin{tabular}{p{.57\columnwidth}p{1.37\columnwidth}}
\toprule

RQ1: Do accuracy and fairness affect men's and women's trust differently? &

Our results indicate that women trusted the fairer model more often than men did, while men tended to prioritize performance. When the model's bias disadvantages their gender, the bias threshold for people to choose the fairer model was lower for women than men. (Section~\ref{sec:menwomenresults}) \\

RQ2: Does making the decision on behalf of a client vs.\ oneself affect trust? &

Our results indicate that participants tolerated more bias when deciding for themselves than when deciding on behalf of a client. (Section~\ref{sec:selfclientresults}) \\

RQ3: Does model performance magnitude affect how much bias affects trust? &

Our results indicate that for models with lower performance, participants trusted the fair model slightly less often, prioritizing performance slightly more. (Section~\ref{sec:modelperformance}) \\

RQ4: Does describing the models' history using textual descriptions and bar charts affect trust? &

Our results indicate that participants trusted the fair model more often when its history was described using text than bar charts. Participants behaved similarly within multiple different textual and graphical representations, including orientation and color. (Sections~\ref{sec:visualrepresentation}~and~\ref{sec:exp2quant}) \\

RQ5: Do demographics and personal characteristics affect people's behavior? &

Our results indicate that willingness to trust, behavioral inhibition and activation scores, and cognitive reflection test scores are all associated with differences in model-choosing behavior. (Section~\ref{sec:demographics}) \\

RQ6: What strategies do people follow in selecting which model to trust? &

We identified seven strategies.
Some participants explicitly quantify and avoid a model's bias, 
while others ignore bias and rely on average performance instead. 
Others still prefer the model that historically preferentially treated others like them. (Section~\ref{sec:Strategies}) \\

RQ7: Does explicitly labeling a model as unfair (whether or not it is) affect trust? &

Our results indicate that participants were less likely to select the model labeled as biased, even if that model was actually more fair. (Section~\ref{sec:explicitbias}) \\

\bottomrule
\end{tabular}
}
\caption{Our study answers seven research questions to understand people's trust in ML models.}
\label{fig:rqs}
\end{figure*}

\looseness-1
Figure~\ref{fig:rqs} summarizes the seven research questions underpinning our experiments and their respective findings. 
Through detailed statistical analyses, we generate psychometric functions describing trade-offs in men's and women's perceived trustworthiness of a model based on its fairness and accuracy, and across visual representations and stakeholder-model relationships. 
We complement our statistical analyses with qualitative analyses of participants' self-reported reasoning strategies.

\pheading{Contributions:} By exploring ML trust and fairness in the context of investment in the presence of gender bias, we synthesize five key insights and contributions of broad relevance to ML fairness visualization and attempt to empower decision-making. 
First, we provide empirical evidence that visualization design choices significantly impact people's prioritization of fairness over performance, influencing trust, as evidenced through detailed comparisons of bar charts and text conditions. 
Second, we demonstrate that men and women weigh accuracy-fairness trade-offs differently when provided with identical visual stimuli. 
Third, we show that an individual's relationship to the model (whether the model's outcome affects the individual or someone else) and explicit warnings of bias can impact trust more significantly than other factors. 
Fourth, we identify a set of strategies people use when reasoning about ML models.
And fifth, we translate our findings into a series of design recommendations for practitioners developing ML fairness visualizations and visual analytics tools.

\section{Related Work}
\label{sec:RelatedWork}

A rich body of research has studied visualization of ML models to support analyses~\cite{HohmanEtAlTVCG2019, YuanEtAlCVM2021, ChatziInfoVis2020}. 
Model fairness has emerged as a particularly important aspect of ML models to visualize~\cite{Cabrera19, Johnson23fairkit, WexlerTVCG2020, AhnLinTVCG2020, WangEtAlTVCG2021, XieEtAlTVCG2022, BhavyaKlausTVCG2023, ZhangEtAlTVCG2023, MunechikaVIS2022}.
Our study empirically explores how design choices and model properties impact a person's trust in ML models, contributing to understanding the design space of ML model visualization.
A broad array of  
visual analytics systems support model fairness assessment and potential remediation of bias. 
Microsoft's Fairlearn~\cite{bird2020fairlearn} and IBM's AI Fairness 360~\cite{Bellamy16} 
implement several fairness metrics and learning algorithms for enforcing fairness and visualizing fairness and accuracy.
Fairkit-learn~\cite{Johnson23fairkit} also visualizes the Pareto optimal frontier of a set of models with respect to model metrics. 
FairSight~\cite{AhnLinTVCG2020}, FairVis~\cite{Cabrera19}, and SliceTeller~\cite{ZhangEtAlTVCG2023} are visual analytics systems that also incorporate model fairness in supporting decision-making. 
The What-If Tool~\cite{WexlerTVCG2020} 
enables non-programmers to visualize datasets and perform counterfactual analysis and observe the effects of data changes on a TensorFlow model. 
Using causal modeling, Discrilens~\cite{WangEtAlTVCG2021} leverages novel set-based visualizations to explore model bias  and 
D-Bias~\cite{BhavyaKlausTVCG2023} interactively supports bias identification and mitigation for tabular datasets. 
FairRankVis~\cite{XieEtAlTVCG2022} supports bias assessment of ranking algorithms.

A user's demographics and computer literacy, the model's actual fairness, the textual description of the model, and the model's transparency and development process can impact user perception of model fairness and trustworthiness~\cite{wang2020}. 
We explore the additional impact of visual design choices for analytic tools for ML, with the goal of improving tool effectiveness.
People's perception of a model's fairness depends on how information is represented. 
For example, too much information can overload participants and result in low quality decisions~\cite{10.1145/3411764.3445315}, and scatterplots can lead to a lower perception of fairness than text~\cite{berkel2021}. 

Visualization research  
uses a variety of trust definitions that are not always validated and can be inconsistent;
using rigorously tested metrics to minimize bias and ensure repeatability can help~\cite{elhamdadi2022trust}.
One of the most common trust measurement approaches 
is asking participants to self-report 
how much they trust a visualization or believe in its accuracy~\cite{xiong2019examining} via a questionnaire or an interview~\cite{7192716, 10.1145/3490099.3511140, 10.1145/3476068}. 
But self-reported measures can be inaccurate and interpretations of scales can vary~\cite{paulhus2007self}.
There are discrepancies between self-reported and behavioral measures of trust~\cite{10.1145/3495013} (the two are weakly correlated), suggesting that they are inherently different~\cite{DANG2020267}.
For these reasons, we employ a trust game~\cite{BERG1995122} to measure users' trust.  
Trust games define trust as occurring when a trustor gives resources to a trustee with no enforceable commitment from the trustee~\cite{coleman1990foundations}. 
For example, a trustor can lend their car to a trustee, knowing the risk that the trustee may not give it back. 
A typical trust game involves two anonymously paired participants: the trustor starts with money, some of which they may choose to give to the trustee. 
The experimenter triples the transferred money, and the trustee can then return some portion back to the trustor~\cite{BERG1995122}. 
Giving more money indicates more trust; while altruistic behavior can also explain giving more money, altruism is typically not the cause of trust-like behavior~\cite{brulhart2012does}.
Trust games can limit the two participants' decisions to a ``trust'' or ``do not trust'' decision, ensuring payoffs reward mutual trust but penalize asymmetric trust~\cite{kreps1990corporate}.
Trust games can take many forms, such as interactive ones with multiple trials testing for cooperation and defection~\cite{komorita2019social, zheng2002trust}.
We use a single round; the trustor selects between two models with which to invest money based on the models' history of returns.

Of the two closest papers to our work, one examined how ML model accuracy affects trust, showing that a difference between stated and observed accuracy reduced trust~\cite{yin2019understanding}. 
The other studied how participants of various races perceived models that discriminate against white and black people, and found that human judges inspired more trust than models~\cite{10.1145/3351095.3372831}.  
Because fairness is domain-specific~\cite{Dwork12, Thomas19}, and many definitions are mutually incompatible~\cite{friedler2016}, we use one common definition, demographic parity, which requires the distributions of model predictions to be similar for the sensitive groups~\cite{Dwork12, Galhotra17fse}. 


\section{Experiment 1: People's Trust in ML Models}
\label{sec:Experiment1}

\looseness-1
We presented participants with pairs of ML models and asked them to select one \model from each pair to invest with. 
We varied characteristics of these models, such as fairness and performance. 
We operationalized performance as the rate of return on investment and fairness as the difference between the return for men and women.
In each pair, one model was generally more fair but had a lower average return than the other.

\subsection{Study Design}
\label{sec:design}

We examined four independent variables: the visual representation (bar chart or textual description), the scenario (invest for oneself or on behalf of a client), average model performance (low or high), and the difference in average returns between the fair and biased model (biased model returns 10\% or 20\% more, on average).
We adopted a mixed-subject design for this experiment and counterbalanced our conditions such that half of the participants saw one level of every between-subject variable.
We describe the manipulations for these variables next. 

\pheading{Visual Representation (between-subject)}:
We used two types of representations to communicate ML model performance: a textual description and a bar chart (orange boxes in Figure~\ref{fig:bartextsptablereps}).

\pheading{Scenario (between-subject)}: We used two scenarios: in one, participants made the investment on their own behalf, and in the other, on behalf of a client whose gender was unspecified.

\pheading{Model Performance (between-subject)}:
We created two conditions of model performance based on the average return on investment from the fair model.
The fair model either had high performance (return on investment of 50\%) or low performance (10\%). 
We chose 50 as substantial enough for the participants to make an informed decision between the fair or biased model, and 10 to allow for a condition where one gender lost money with the biased model, which returned less money than it was given.

\pheading{Difference in Average Return (within-subject)}:
For each tier of model performance (high vs.\ low), we manipulated the differences between the average return on investment for the fair and biased model to be either small (10\%) or large (20\%). 
For example, for the small-difference condition, the high-performing fair model returned 50\% to both men and women, while the competing biased model returned 40\% to men and 80\% to women, so its average rate of return of 60\% is 10\% higher than the fair model.
We adopted a pseudo-staircasing method to generate the specific return values for the biased models. 
In perception research, staircasing methods involve increasing or decreasing the discriminability of a presented stimulus depending on the participants' response~\cite{rensink2010perception, otto2009comparative}.
Staircasing allows researchers to identify the just-noticeable difference between two intensities of a stimulus.
In our study, we pit a fair model against a biased model that has a higher average return to observe the threshold for people to be willing to choose the biased model despite its bias.
We vary the model's degree of bias by increasing or decreasing the differences in return to men and women.
Tables~1--4 in the SM~\cite{SM} show all the conditions we tested.

For all the experiments, participants completed several psychometric tests to evaluate the potential impacts of individual cognitive and personality differences on model selection:

\pheading{Cognitive Reflection Test (CRT)} contains three quantitative questions shown to be correlated with quantitative reasoning ability~\cite{frederick2005cognitive}.
A good CRT performance suggests that the participant took the survey seriously and possesses decent quantitative reasoning skills.

\pheading{Rotter's Interpersonal Trust Inventory} survey~\cite{rotter1980interpersonal} consists of 25 statements 
that measure an individual's tendency to trust others. 

\pheading{Behavioral Inhibition System and Behavioral Activation System (BIS/BAS)}
measure sensitivity to punishment/reward and motivation to inhibit behavior that results in negative outcomes/encourage seeking the achievement of goals~\cite{bisbas}.

\subsection{Procedure}

\begin{figure}[t]
    \centering
    \includegraphics[width=\columnwidth]{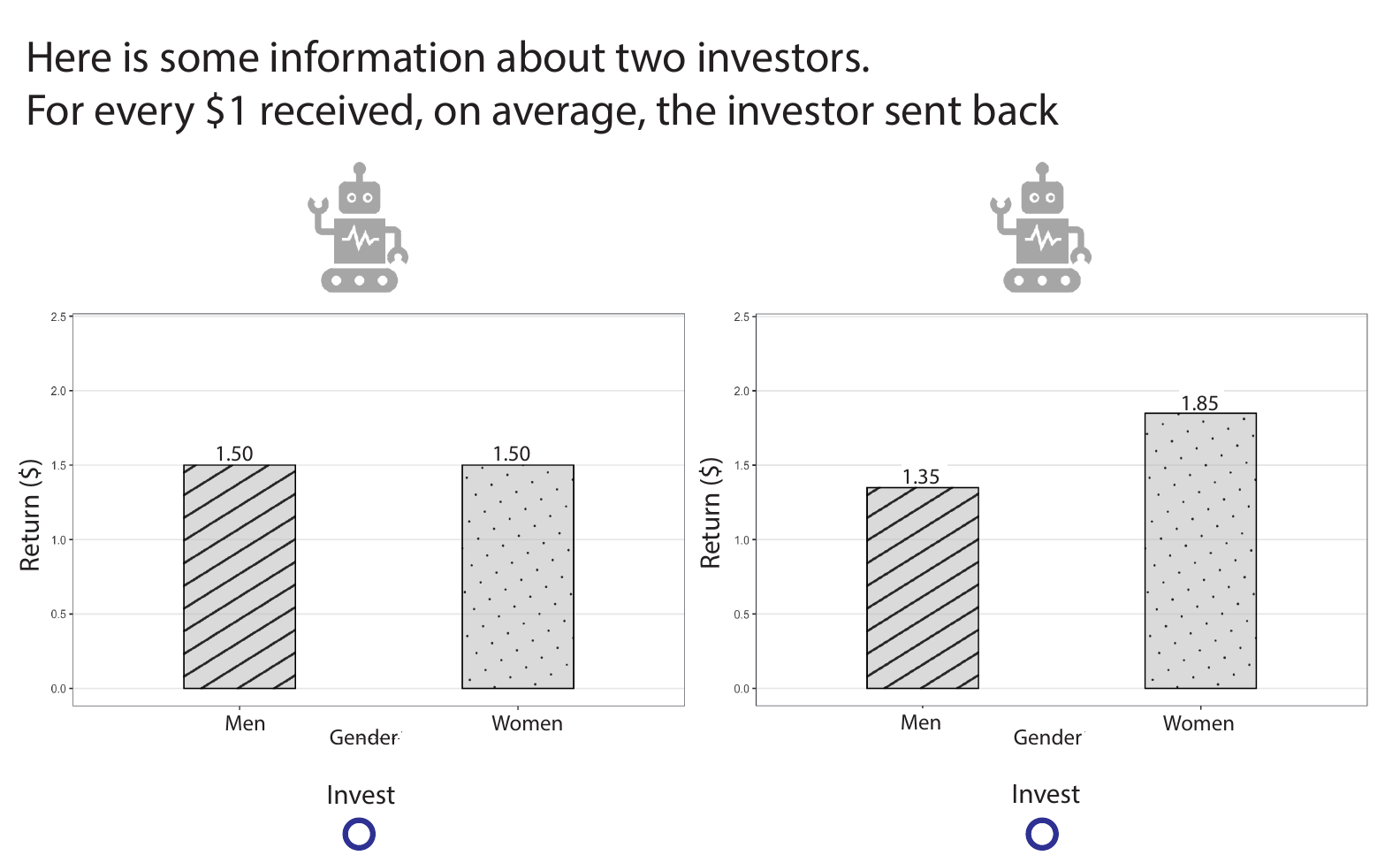}
    \caption{An example question using the bar chart representation.}
    \label{fig:barquestionexample}
\end{figure}

We deployed the study using Qualtrics~\cite{qualtrics2013qualtrics} and distributed it via Prolific.co~\cite{palan2018prolific}. 
The survey started with a consent form.
Next, the participants completed the Cognitive Reflection Test (CRT) and the Interpersonal Trust Inventory. 
They then read a brief introduction to the survey and performed several rounds of a sample of our investment trust game.
They were told each model's rates of return depended on the model's accuracy: 
The more accurate the model was, the higher average returns on investment it produced. One of the models provided the same rates of return on investment for both men and women.
The other model was biased and provided a higher return rate to one gender, but its average return rate was higher than that of the fair model. 
We counterbalanced the gender towards which the model is biased, such that half of the time the model was unfair to men, and the other half to women. 
We told the participants, ``Before you choose which investor you want to invest with, you will see some information about how these investors have performed in the past''.
We neither explicitly informed participants of the model's average return nor bias, but showed the returns for men and women for each model in either text or bar chart form. 
Figure~\ref{fig:barquestionexample} shows a bar-chart version example.

Next, participants completed 48 rounds of the trust game, covering the conditions outlined in Section~\ref{sec:design}.
The rounds were presented in a random order. 
We recorded participants' investment choices for each pair of models. 
We also provided participants a free-response text box to explain their reasoning for 16 of the rounds, randomly distributed throughout the survey. 
See Section~\ref{sec:Strategies} for more details.

We included six attention checks throughout the survey.
One attention check was related to their visual reasoning skills (look at a bar chart and select the tallest bar).
The second tested their basic mathematical reasoning skills (how much money would they get in return if they invest \$10 and the model returns \$1.50 for every \$1). 
The other four attention checks showed participants pairs of fair models, one providing a higher return than the other;
to pass, the participants needed to select the higher-return model. 
We excluded participants who failed at least one attention check from our analysis. 
At the end of the survey, participants completed the BIS/BAS inventory and reported their demographic information. 
Finally, participants reported how much effort they put into completing the study.
They were assured that the answer to this question would not affect their compensation and were encouraged to answer honestly. 

\subsection{Hypotheses and Expected Outcomes}
\label{sec:exp1_expected}

\begin{figure*}[t!]
    \centering
    \includegraphics[width=\textwidth]{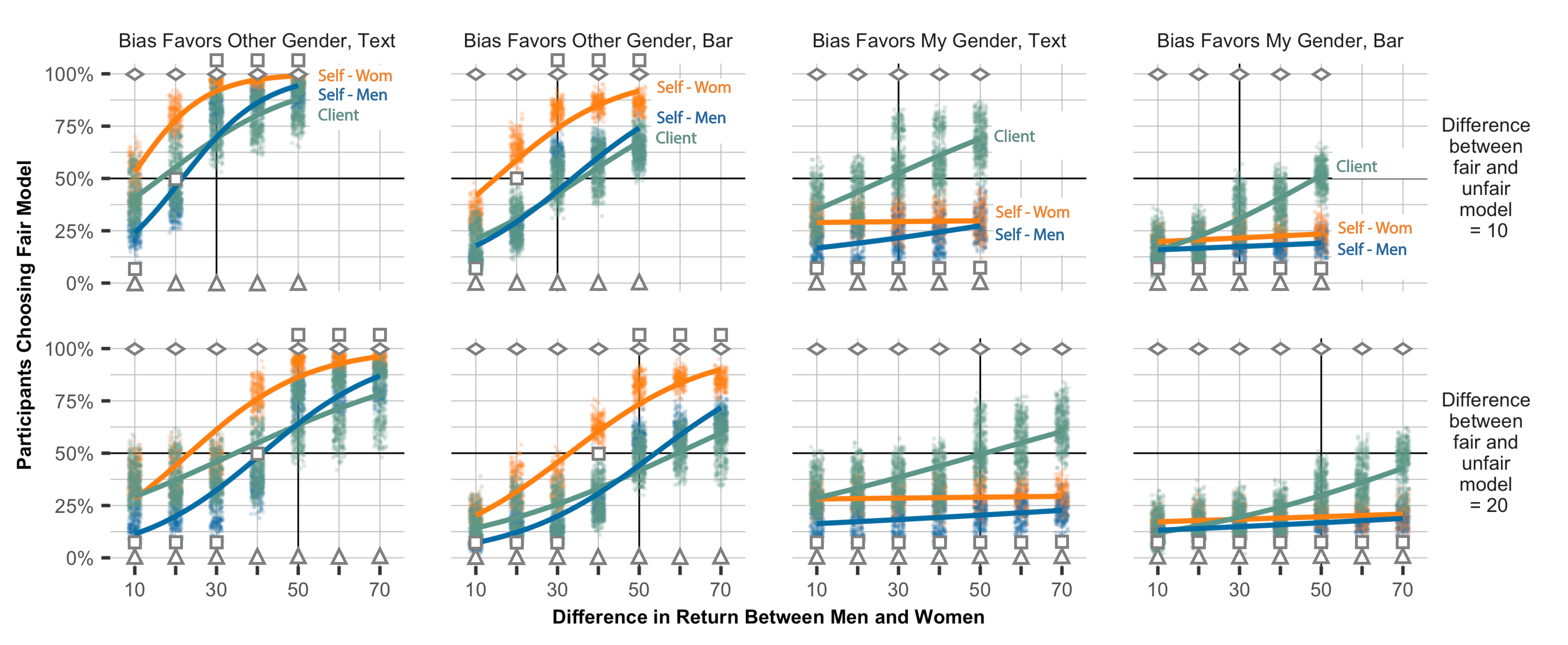}
    \caption{Mean subset plots and logistic regression lines for bootstrapped results data. Data where participants are investing on behalf of a client is labeled ``Client'' (teal), and data where they are investing on their own behalf is separated by gender and labelled ``Self~-~Men'' (blue) and ``Self~-~Wom'' (orange). The X-axis represents the difference between returns to men and women by the biased model, and the Y-axis values represent the percentage of participants who chose the fair model for each bootstrapped data set. Subplots are separated by representation type (bar vs.\ text) along with whether the biased model returns more to the participant's gender or the other gender. The square, triangle, and diamond represent the three possible extreme behaviors that participant might exhibit: gender-aware, maximizing profit; gender-blind, maximizing profit; and maximizing fairness, respectively (recall Section~\ref{sec:exp1_expected}).}
    \label{fig:exp1_self_client}
\end{figure*}

To contextualize possible participant behavior, we discuss the rational behavior that three possible extreme participants might exhibit. The ``gender-aware, maximizing profit'' participant always selects the model that has historically produced the highest return for members of the participant's own gender.  This participant is unaffected by a model's potential bias.  The ``gender-blind, maximizing profit'' participant also ignores bias but selects the model that historically maximized the average returns for all past clients, averaging the historical returns for women and for men. This strategy might particularly make sense when the participant does not know the gender of the client on whose behalf the investment is being made. Finally, the ``maximizing fairness'' participant always selects the model that minimizes the difference in historical returns for women and for men, ignoring the overall historical profit the models have produced. Figures~\ref{fig:exp1_self_client},~\ref{fig:bartextsptablereps},~and~\ref{fig:barexplicit} indicate how each of the three extreme participants would perform, and how the actual participants' behavior compares to theirs.

Furthermore, existing work has shown that explicitly showing numbers facilitates mental computation of differences between values, while bar charts direct readers' attention to salient, large values~\cite{xiong2022reasoning}.
Thus we hypothesize that textual representation will elicit more fair behaviors as participants are more likely to notice the difference between returns for men and women, while bar charts will elicit less fair behaviors as they draw attention to values associated with larger profits. 
We expect visual representation to interact with the underlying data, which represents model performance, as larger data values tend to translate to more salient differences. 
We also expect to see significant individual differences as data can be personal~\cite{peck2019data}.
We hypothesize that since the model depicts gender-related data, there will be systematic differences in investment behaviors between men and women.

\subsection{Participants}

We recruited 1,347 participants for this experiment using Prolific.co~\cite{palan2018prolific}. 
Participants were compensated at \$12 USD per hour. 
We filtered for participants who were fluent in English, over 18 years old, and reside in the United States. 
After filtering responses to remove attention check failures and nonsense, we were left with 1,326 participants (599 women, 608 men, 119 non-binary or indicated to prefer to self-describe, $M_{\textit{age}}$ = 37.0, $SD_{\textit{age}}$ = 13.2).

\subsection{General Analysis Approach}
\label{sec:exp1-results} 

We used R for all statistical analyses. 
We made our data and R scripts public: \url{https://osf.io/er5a3/?view_only=cda4c6acfd684da287225c8124fb7b9e} (hereon referred to as SM)~\cite{SM}. 
We constructed two models to find statistically significant effects of our studied variables and their interactions: 

\pheading{Logistic Regression:} We combined data from all studies from Section~\ref{sec:design}, filtered for responses from those identifying as men and those identifying as women to build this regression model. 
The dependent variable predicted by the  regression model was the participant's choice to use the fair~(1) or the unfair~(0) model. 
The linear predictor formula included a set of predictors needed to answer research questions RQ1--RQ4, as well as all of their second-order interactions --- gender, the direction of bias (consistent with participant gender or not), scenario (self or client), the return of the fair model and the maximum difference between returns for the biased model, 
and representation type (bar chart or text). 
We also included CRT scores, trust scores, BIS/BAS scores, and all collected demographics, including age, income, race/ethnicity, and education. 
This model also included second-order interactions between gender and scores, as well as gender and demographics. 
We created a type-II analysis-of-variance (ANOVA) table for this model, and report the $\chi^2$ values and associated $p$ values for statistically significant main effects and second-order interactions in the sections below. 
We also conducted a test for multicollinearity for all the co-variates and found them to be largely independent, with variance inflation factors between 1.00 and 1.63, well below the suggested cutoff of 10~\cite{multivariateanalysis}.

\pheading{Bootstrap Re-Sampling:} To approximate a measure of uncertainty in the data, we performed bootstrap re-sampling and fit a linear model to the resulting data separately, for those identifying as men and women. 
For each gender, we took $N$ samples with replacement, where $N$ was the number of respondents with that gender.
We did this 100 times and aggregated, by taking the mean of the results across each of the above predictors, excluding all scores and demographics aside from gender. This gave us a percentage of participants who chose ``fair'' for each combination of predictor values.
This aggregation provided us with percentage values of participants choosing the fair \model, which we then used as the dependent variable for the fitted linear model. 
The formula included the main effects and all second-order interactions of the variables across which we aggregated. 
We used this model to calculate the estimated marginal means of the dependent variable for each of the significant predictors in the regression model. An analysis-of-variance table with effect sizes can be found in the SM~\cite{SM}.

\subsection{RQ1: Effects on Men's and Women's Trust}
\label{sec:menwomenresults}
Figure~\ref{fig:exp1_self_client} shows how men and women behaved when investing on their own behalf using text and bar chart model representations. 
On average, everyone was more likely to choose the model that gave their gender the higher return, which tends to be the biased model. 
However, women chose the fair model about 1.5 times more often than men  $\chi^2 = 480.92$, $p < 0.001$). 
Overall, $48.7\%$ of women chose the fair model, while $35.9\%$ of men did ($SE = 1.25 \times 10^{-3}$).

When participants invested on their own behalf and the biased model favored their own gender, 
17.3\% of men and 26.1\% of women chose the fair model.  
When the model was biased against their own gender, 49.6\% of men and 68.5\% of women chose the fair model. 
This pattern continued when participants chose on behalf of a client.
When the biased model favored the participants' gender, 35.0\% of men and 41.6\% of women chose the fair model. 
When the bias was against, 41.7\% of men and 58.4\% of women did so.

We identified another asymmetry between women's and men's behavior. 
Recall that the biased model always generated higher average returns, so there are trials where a gender receives the same return from both the fair and the biased models, while the other gender gets an even higher return from the biased model.
In these scenarios (e.g., fair: 50\% to men and women, biased: 70\% to men and 50\% to women), men (69.3\%) were more likely to choose the biased model than women (41.4\%), even if their own gender was being discriminated against.

\subsection{RQ2: Effect of Choosing For Yourself vs.\ a Client}
\label{sec:selfclientresults}

We compare how trust in ML models changes when participants invest not for themselves, but on behalf of a client of an unspecified gender. 
We refer to these two conditions as two investment scenarios. 

Participants were on average 3.25 times more likely to choose the fair model when investing on behalf of a client than themselves ($\chi^2=74.34$, $p < 0.001$). 
They chose the fair model more often on clients' behalf ($38.3\%$) than on their own behalf ($21.7\%$) 
when their gender received a higher return, but less often on clients' behalf ($50.1\%$) than on their own behalf ($59.0\%$) when their gender received a lower return.
But their gender significantly interacted with their tendency to choose the fair model depending on the scenario ($\chi^2=18.21$, $p < 0.001$).
Overall, $33.4\%$ of men and $47.3\%$ of women participants choosing on their own behalf chose the fair model, while $38.4\%$ of men and $50.0\%$ of women choosing on behalf of a client did so. 
Participants also became more likely to choose the fair model on behalf of a client as the bias of the biased model increased (Figure~\ref{fig:exp1_self_client}).

Additionally, in conditions where the fair model gives one gender the same amount as the biased \model, participants were more likely to choose the biased one.
Filtering for these conditions, estimated marginal means show that 33.9\% of participants chose the fair model.

\subsection{RQ3: Effect of Model Performance}
\label{sec:modelperformance}

We next considered how fairness-performance behavior trade-off changed when we varied the baseline performance of the fair model between high performance (50\% return) and low performance (10\% return).
Participants were 1.41 times more likely to choose the fair model when the average model performance was 50\% compared to 10\% ($\chi^2=30.07$, $p < 0.001$).
Participants are 1.29 times as likely to choose the fair model when the difference between the fair and biased model is smaller (i.e., a difference of 10\% vs.\ 20\% in average returns) ($\chi^2=310.20$, $p < 0.001$).
Figure~\ref{fig:exp1_self_client} shows that when participants are disadvantaged by the biased \model and choosing on their own behalf, they tend to choose the fair model at least 50\% of the time once the fair \model begins to offer the same or a higher return than the biased model. 
When the bias is advantageous, both men and women have no tipping point when investing for themselves. 
When choosing on behalf of a client, there is a more gradual increase in participants choosing the fair \model as the discrepancy between returns grows. 

\subsection{RQ4: Effect of Visual Representation}
\label{sec:visualrepresentation}

We next investigated the effect of visual representation of model information on fairness perception and trust.
Figure~\ref{fig:exp1_self_client} shows the main effect of visual presentation ($\chi^2=914.23$):
participants were 1.48 times more likely to choose the fair model when the model information was presented as text (49.8\%) compared to bar charts (34.8\%). 
We also found a significant interaction between visual representation and model bias ($\chi^2=11.00$).
Although overall participants were more likely to choose the fair model when their own gender was being discriminated against (and vice versa), the tendency to choose the fair model was stronger in the text ($63.3\%$) condition, compared to the bar chart condition ($45.8\%$). 
This remains true when participants' own gender was being favored by the biased model.

\subsection{RQ5: Demographics and Personal Characteristics}
\label{sec:demographics}

We next assessed whether different participant demographics, along with personal characteristics such as trust scores, are correlated with different decision-making patterns.
We found a main effect of trust scores: participants were more likely to choose the fair model ($\chi^2=49.22$, $\textit{Odds Ratio(OR)} = 1.01$) if they scored higher on the trust inventory. 
Participants were also more likely to choose the fair model when they scored higher on the BIS ($\chi^2=82.24$, $OR= 1.01$) and BAS drive ($\chi^2=10.60$, $OR=1.03$)
inventories, but less likely when they scored higher on BAS reward ($\chi^2=13.82$, $OR=0.99$). 
For CRT, participants who scored higher CRT scores were less likely to pick the fair model ($\chi^2=447.05$). 
We also found an effect of age, education, level of income, and race/ethnicity, but the effect sizes were negligible. Details can be found in the SM~\cite{SM}. 

\begin{figure*}[t!]
    \centering
    \includegraphics[width=\textwidth]{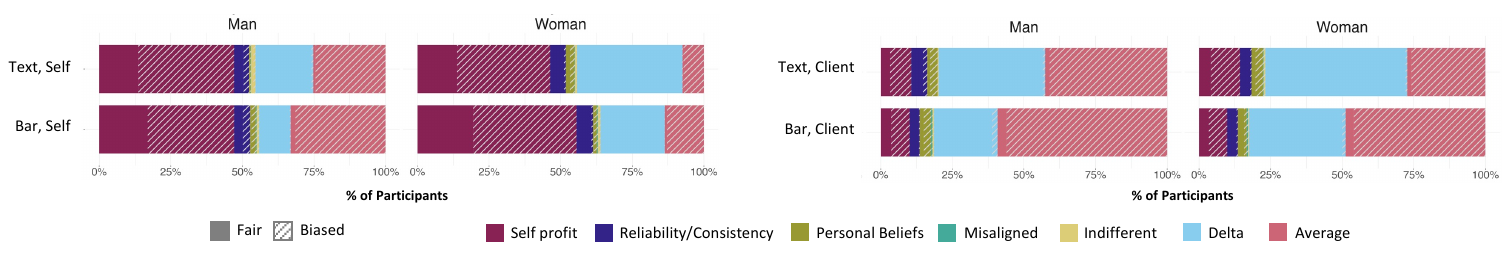}
    \caption{The fraction of participants using each strategy who chose the fair or biased \model in Experiment~1 (left) and Experiment~3 (right).}
    \label{fig:codingstrategies}
\end{figure*}

\subsection{RQ6: Reasoning Strategies}
\label{sec:Strategies}

To understand the rationale behind participants' choices in the trust game and how they interpreted the history of return information, we conducted an inductive thematic analysis~\cite{doi:10.1191/1478088706qp063oa, inbook} on the reasoning they provided for selecting a \model in the free-response questions.
Two authors independently constructed a set of codes from a subset of the data after going through each response. 
They then conducted a converging exercise to integrate their codes into a standard set of codes, with definitions and prototypical examples. 
The two authors then used these codes to categorize all participant responses, with each author independently responsible for half of the coding. The first authors reviewed each categorization for consistency with code definitions.

\subsubsection{Strategies}

We identified seven strategies the participants used (see Figure~\ref{fig:codingstrategies}): 

\pheading{Average}: The participants computed the average return for each \model and compared the two \model's average returns, leveraging their computation to either maximize profit or fairness.

\pheading{Delta}: The participants computed the model's bias (difference in return between the two genders) and compared the two models' biases. 

\pheading{Indifferent}: The participants were indifferent about their choice.

\pheading{Misaligned}: The participants' response did not align with the choice that they made, such as selecting the more biased \model, having explained as selecting it for being less biased. 

\pheading{Personal Beliefs}: Participants used information that was not provided, such as by making up assumptions for why a \model favors one gender. 

\pheading{Reliability/Consistency}: The participants reported that they selected the model that was ``safe,'' ``reliable,'' or ``consistent across trials.''

\pheading{Self Profit}: The participants chose the model that historically had higher returns for their gender. This strategy corresponds to the ``gender-aware, maximizing profit'' type as described in Section~\ref{sec:exp1_expected}.

\subsubsection{Results}
\label{sec:rq7results}

As shown in Figure~\ref{fig:codingstrategies}, women tended to use the delta strategy more than men and chose the fairer model more often.
The delta strategy involves computing model bias by calculating the discrepancies between returns in the two models.
The affordance of text on difference computation might be the driving factor behind participants more frequently choosing the fair model. 
This inference aligns with existing work on how people reason with data:
explicitly showing numbers facilitates difference computations, while visualizing values with bar charts draws people's attention to salient large values instead~\cite{xiong2022reasoning}.
This bottom-up attraction to the salient large bars, along with the top-down effects of paying attention to self-relevant data, potentially explains why participants more often used the self-profit strategy with bar charts.
Depending on whether that self-relevant bar happened to be relatively large or small, participants ended up choosing the biased or fair model.
We see supporting evidence of this, as the self-profit strategy is less predictive of selecting the fair model than the delta strategy.
This effect of letting the salient large bar and self-interest drive attention and decision became attenuated when the participants made decisions on behalf of a client, as the participant was left without having a specific gender bar to focus on.
Overall, fairness perception appears to be driven by the perceptual salience of self-relevant data values (which corroborates existing findings that suggest data is personal~\cite{peck2019data}), and how much the representation affords difference computation. 
Ultimately, fairness perception in visualizations seems closely related to the ease of perceiving differences between groups, potentially indicating that designing visualizations to highlight between-group differences and minimizing the salience of one large value related to self-interest might sway people from choosing the biased models.

\subsection{Non-Binary Participant Data}
\label{nonbinaryResults}

We received 119 responses from participants who did not identify as men or women.
Among them, 93 participated in the scenario where they invested on their own behalf and the average return was 50\% for the fair model, and 60\% or 70\% for the biased model. 
We share some anecdotal results on this limited dataset to provide preliminary insights into how non men and women reacted to our experimental set-up.
Future work should more closely and systematically examine these effects for more in-depth insights.

We performed the same bootstrap re-sampling as above for this condition set but included those identifying as non-binary, who preferred to self-describe, or who preferred not to disclose. 
The direction of bias was unclear to this group, as the model information displayed in the trust game does not include a history of performance for non men and women. 
The only linear predictors were gender and maximum return difference. 
Both gender ($\chi^2 = 254.45$, $p < 0.001$, OR $\frac{\textit{women}}{\textit{men}} = 1.78$, OR $\frac{\textit{non-binary}}{\textit{men}} = 2.67$) and maximum return difference ($\chi^2 = 55.54$, $p < 0.001$, OR $\frac{70}{50} = 0.694$) had significant main effects (but no significant interaction). 
We found a significant difference between how participants who identified as non-binary interacted with participants who identified as men or women.
Recall that the fair model in this batch of data always returned a profit of 50\%. 
Of non-binary participants, 64.6\% chose the fair \model, more often than women~(56.0\%) and men~(41.5\%). 
However, when filtering the men's and women's responses to those where the gender doing the choosing was not advantaged by the bias, women chose the fair model most often~(78.8\%), followed by non-binary participants~(64.6\%), and men~(56.4\%). 

Qualitatively, non-binary participants reported sometimes being indifferent to their choices, as they did not identify with either of the genders that received bias benefits.
Some non-binary participants used their assigned birth sex to make decisions. 
Others mentioned being against sexism in general. 
We further discuss these topics in Section~\ref{sec:ImplicationsandFW} and share our insights on how to better account for their experiences and capture their responses in visualization and related research. 

\begin{figure*}[t!]
    \centering
    \includegraphics[width=\textwidth]{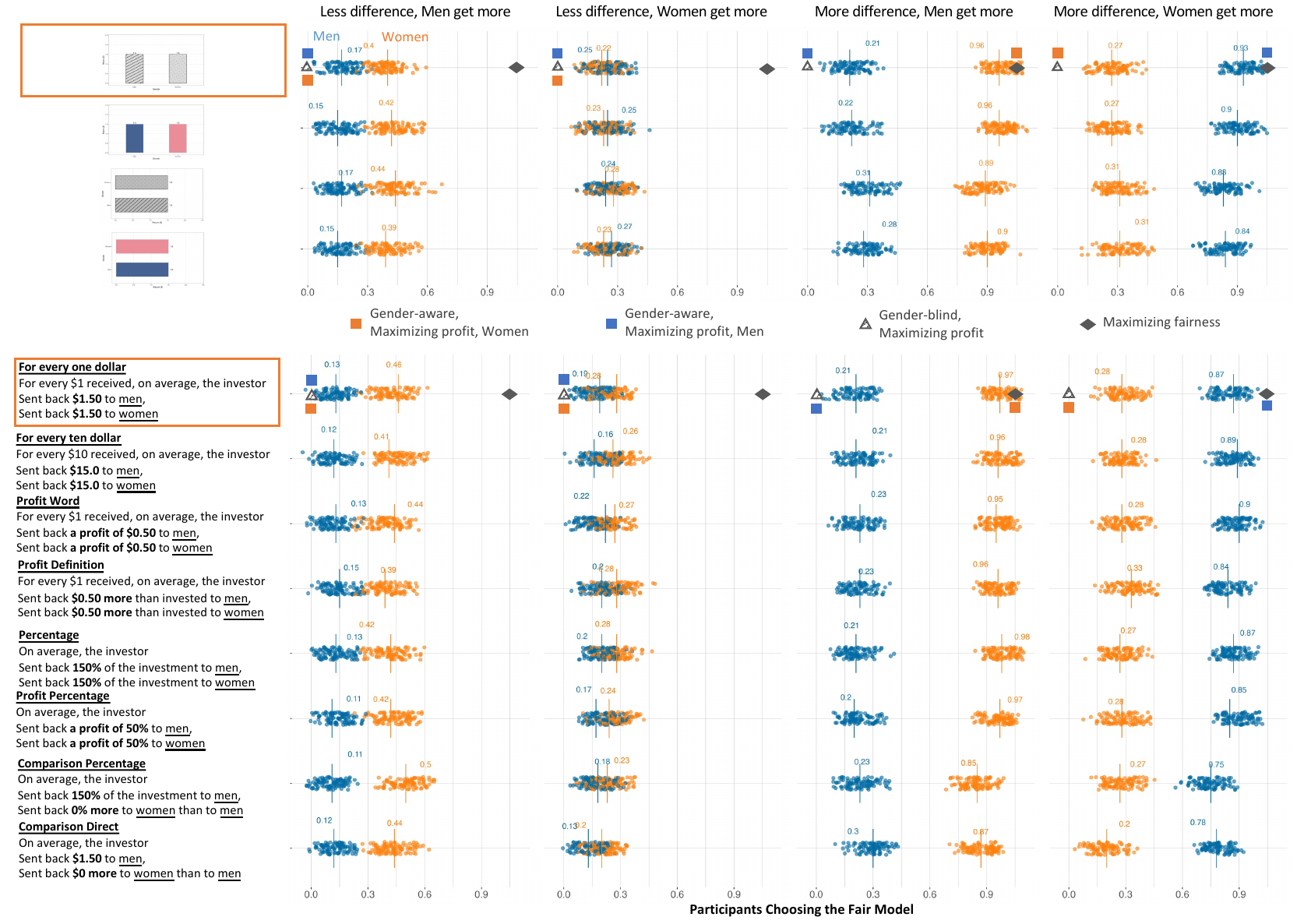}
    \caption{The results of Experiment~2, along with the style of the 8~textual, and 4~bar chart  representations. The visualizations covered in orange box were used in Experiment~1. The orange dots show the percentage of women who choose the fair model, and the blue dots show the percentage of men who did so. The square, triangle, and diamond represent the three possible extreme behaviors that a participant might exhibit: gender-aware, maximizing profit; gender-blind, maximizing profit; and maximizing fairness, respectively (recall Section~\ref{sec:exp1_expected}).}
    \label{fig:bartextsptablereps}
\end{figure*}

\subsection{Discussion and Summary}
\label{discussionexp1}

Women were more likely to choose the fair model than men. Analysis of reasoning strategies (Section~\ref{sec:Strategies}) suggested that this might be driven by women more focused on comparing the differences between the two models and looking at the degree of bias, and men focused on the absolute value of the model with the higher return. 

Participants, on average, more often used the `self-profit' strategy when they invested on their own behalf and chose the less fair \model. 
When they invested on behalf of a client, they tended to adopt the 'delta' and 'average' strategies and more often chose the more fair \model. They were more likely to prioritize fairness when the overall returns for both model choices were higher. This suggests that people care about the overall model performance and are willing to tolerate more bias if the returns are low. When their gender is disadvantaged, participants chose the fair \model more often if it offered them the same or a higher return. When their gender is advantaged, the tendency to choose the fair model was less likely.
When choosing on behalf of a client, there was a gradual increase in participants choosing the fair \model as the magnitude of the bias increased. 
This indicates that many participants have a tipping point where they begin to prioritize fairness over average model performance when considering the returns to others, but do not acknowledge this trade-off when choosing for themselves.

We found that higher trust score, BIS/BAS score were associated with an increased tendency to choose the fair \model, while higher BAS reward score and CRT score were associated with a decreased tendency to choose the fair \model. This indicates that tendency to trust, sensitivity to punishment, sensitivity to reward, and cognitive reasoning ability all may have some impact on people's tendency to trust \model{s}.

Finally, preliminary findings show that non-men and women participants were more likely to choose the fair \model than men in every condition, and were more likely to choose the fair \model than women only in the case where women were the advantaged gender. However, we cannot make any definite conclusions because those who do not identify as men or women may be assigned at birth as one of these genders, or identify more closely with either one of these genders.


\section{Experiment 2: Style and Phrasing Generalizability}
\label{sec:Experiment2}

Results from Experiment~1 (Section~\ref{sec:visualrepresentation}) suggest that people chose the fairer \model when the \model information was presented as text rather than a bar chart, which joins recent explorations that demonstrated presenting the same data in different visual formats alters the perception of algorithmic fairness~\cite{berkel2021}. 
Next, we test the robustness of our observations on bar charts and textual descriptions across multiple visual styles of bar charts and alternative phrasing of textual descriptions. 
This also enables us to make a fairer comparison between the overall effect of bar charts and textual descriptions. 

\subsection{Hypotheses and Expected Outcomes}
\label{sec:exp2_expected} 

Prior work~\cite{9623314} has demonstrated that semantic associations, including those related to discrimination and sensitive information, impact color selection when constructing visualizations. 
We were concerned that the inverse might hold true, specifically that visualizing bias using colors with known semantic association might alter participants' decision-making~\cite{schloss2020semantic, lin2013selecting}. 
To explore this possibility, we designed this experiment with the returns for men and women visualized using blue and pink, respectively (a traditional North American color convention for gender associated with childhood and adolescence).
We hypothesize that such color changes will have an effect on biasing participants, as the colors might cause them to focus on gender differences.
But the reverse might also hold true.
Participants might be minimally impacted by this color change, as existing work has demonstrated that behavioral biases tend to be robust across color manipulations~\cite{xiong2023whatdoes, xiong2022investigating}.
To further cover generalizability, we test the effect of another visual style change to be compared with the effect of color changes, while also examining the extent to which the fairness bias we observed from Experiment~1 generalizes across bar chart styles.
We manipulate the arrangement by flipping the bars horizontally, inspired by prior work that shows spatial arrangement can change the target, speed, and accuracy of visual comparisons~\cite{ondov2018face, xiong2021visual}.
Existing work has shown that describing probabilities using natural frequencies (e.g., 1 out of 4) tends to promote better numerical understanding and statistical inferences compared to proportions or percentages 
(e.g., 25\%)~\cite{hoffrage2000communicating, fernandes2018uncertainty}.
Therefore, we also hypothesize that variations in text phrasing will have an effect on participants' behaviors, such that frequency-based descriptions will help highlight discrepancies between model returns for the two genders and motivate participants to choose the fair model more often.

\subsection{Design and Procedure}
\label{sec:Exp2-design}

Instead of creating psychometric functions modeling participant behaviors for each bar style and text phrasing, and to keep the length of the experiment manageable, we sampled two levels of return values (counterbalanced for men and women) from the full staircase in Experiment~1 and generated bar charts and textual descriptions using values from these four levels.
We tested 8~textual and 4~bar chart representations (including the ones used in Experiment~1) --- see Figure~\ref{fig:bartextsptablereps}. 
We detail our rationale for selecting the alternative textual descriptions in the SM~\cite{SM}. 
At each level, we compare the bar representations to each other, and the text phrasings to each other. 
We chose two levels that had the biased model either being a little biased (giving one gender 55\% and another gender 65\% return on investment), or extremely biased (giving one gender 35\% and the other 85\% return), with the gender to which the model is biased counterbalanced, totaling four conditions.
The return from the fair model was kept constant at 50 for all conditions. 

\subsection{Participants}
We recruited 413 participants from Prolific.co~\cite{palan2018prolific}.
After applying the same exclusion criteria as in Experiment~1, we were left with 410~participants (195~women, 209~men, $M_{age} = 36.78$, $SD_{age} = 13.14$).
Half were assigned to read the text variations, and the other half were assigned to read bar variations. 
Although we exclusively recruited men and women, we ended up with some participants who identified as non-binary ($N = 4$)
and some selected `prefer to not disclose' ($N = 2$). 

\subsection{Quantitative Analysis}
\label{sec:exp2quant}

We performed bootstrap re-sampling for this data with the same approach as before.
For each bootstrap sample, we calculated the percentage of participants in the sample who chose the fair model for each of the four conditions outlined in Section~\ref{sec:Exp2-design}, across 8 text and 4 bar chart representations.
We performed an ANOVA test on the bootstrapped samples, comparing the percentage of people that chose the fair model across visual representation types (text or bar), and gender.

We found a main effect of visual representation ($F(1, 9504) = 451.190$, $p < 0.001$). 
Participants were more likely to choose the fair \model with bar charts ($42.71\%$ chose fair, $SE = 0.07$) than text descriptions ($40.95\%$ chose fair, $SE = 0.05$). 
For bar charts (see Figure~\ref{fig:bartextsptablereps}), participants behaved similarly across design styles that varied in color pallet and orientation. 
For the text descriptions, participants also behaved similarly across most alternative phrasings.

We also found an effect of the unfair conditions (small vs.\ large discrimination) on fairness perception and choice. 
Participants noticed the trade-offs between model performance and fairness ($F(3, 9504) = 53663.421$, $p < 0.001$).
They preferred the fair model compared to the biased model that more drastically discriminated against one gender, despite that biased model generating a higher average return (e.g., 35\% returned to one gender and 85\% to another).
In these conditions, 57.95\% of the participants chose the fair model. Participants were more tolerant of the biased model that generated a higher return without drastically discriminating against one gender (e.g., 55\% returned to one gender, and 65\% returned to the other). 
On average, only 25.70\% of the participants chose the fair model in these scenarios. 

We observed a similar effect of gender as we did in Experiment~1
($p < 0.001$). Men were less likely to choose the fair model overall ($36.52\%$, $SE = 0.058$), compared to women ($47.14\%$, $SE = 0.058$).
We also saw an interaction between fairness conditions (e.g., 55\%/65\%, 35\%/85\%) and gender ($F(3, 9504) = 107500.732$, $p < 0.001$).
Overall, participants preferred choosing the fair model when the biased model discriminated against their own gender, as shown in Figure~\ref{fig:bartextsptablereps}, where the men and women data flipped between columns. 
Women seemed less willing to choose the biased model than men both in the case where bias was advantageous to their gender and when it was not.

The few participants who identified as non-binary or other chose differently for different conditions. 
They behaved similarly across different bar chart styles, where $31.2\%$ chose to be fair on average. However, we observed that they tended to trust the fair model in the conditions where men get more especially if the difference in return between men and women was large (i.e., men return: 85\%, women: 35\%) where $66.7\%$ chose fair, and for the $65/55$ condition (men return: 65\%, women: 55\%) $29.2\%$ chose to be fair. Whereas in the conditions where women get more (e.g., men return: 35\%, women: 85\%), they tended to trust the biased model more often ($14.6\%$ chose fair). 

\subsection{Discussion and Summary}
\label{sec:exp2discussion}

Contrary to our hypotheses, there seemed to be minimal differences in participants' reactions across bar styles and text phrasings. This suggests generalizability between different variations of bar charts and text representations in impacting selection behaviors.  
Similar to Experiment~1, we found that 
whether overall performance or fairness is preferred depended on the difference in return.
Women seemed to be less willing to choose the biased model, even if it favors women. 
But men seemed to be more willing to choose the biased model when it favors their gender.
However, men seem to also be more willing than women to choose the biased model that is biased against their own gender, which might be due to the social desirability bias~\cite{stark2019impact, sears2005over} as they refrained from expressing potential prejudices against women.

Across the two levels of returns we tested, we also found participants tend to be slightly more fair for bars as compared to text.
This seemingly depicts a contradiction to our results from Experiment~1. 
But this isn't the case. 
It is not valid to compare Experiment~1, which adopted a full staircase design across multiple levels, and Experiment~2, which looked at only two (counterbalanced) slices of that staircase. 
The psychometric function from Experiment~1 compares the summary behaviors between men and women across all levels of fairness when presented with bars and text, and comparing such continuous psychometric function to sampled data at two points can lead to misinterpretations~\cite{goldstein2022leveling}.


\begin{figure*}[t!]
    \centering
    \includegraphics[width=\textwidth]{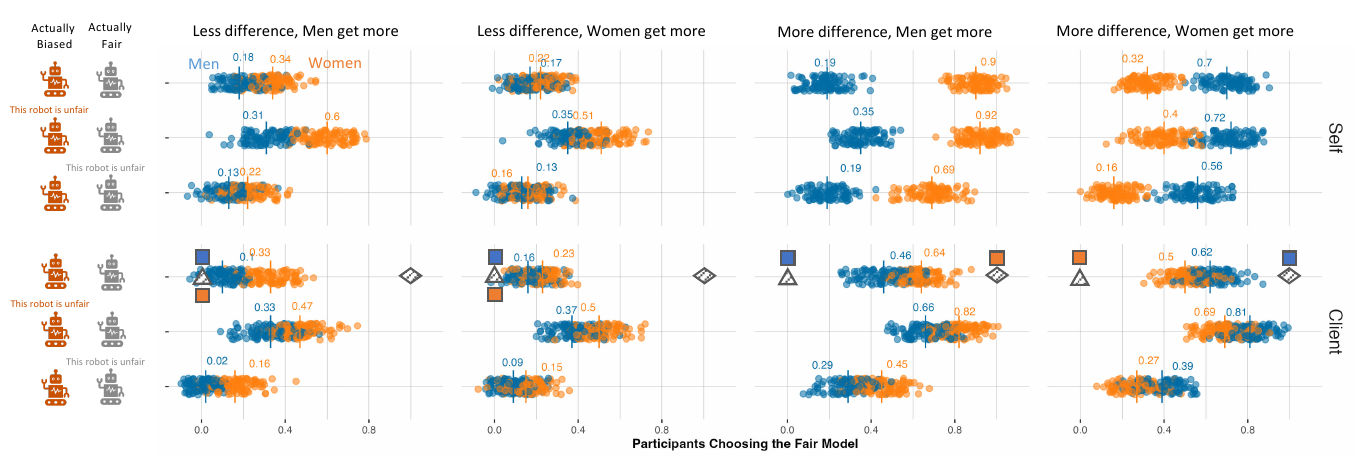}
    \caption{The results showing how people perceive bias when coming across pairs of models, represented using regular bar charts, and bar charts with or without explicit warning (saying it is biased towards a specific gender). The dots in orange show the percentage of women choosing the fair model, and the dots in blue represent men choosing the fair model. The square, triangle, and diamond represent the three possible extreme behaviors that a participant might exhibit: gender-aware, maximizing profit; gender-blind, maximizing profit; and maximizing fairness, respectively (see Section~\ref{sec:exp1_expected}).}
    \label{fig:barexplicit}
\end{figure*}

\section{Experiment 3: Effect of Textual Annotations}
\label{sec:Experiment3}

Experiment~1 showed that participants were more likely to choose the fair model when the model information was presented as text compared to bar charts, and Experiment~2 suggested that the visual styles of the bars and phrasing of the text had negligible effects on participants' behaviors beyond the effect of gender, model performance, and investment context (self vs. client).
However, in the real world, data is often presented leveraging both visual and textual annotations.
In this experiment, we explore the interactive effect of text and bar charts on fairness perception and model selection.
We focus on textual annotations, which prior literature has illustrated to profoundly impact user preference and data comprehension and preferences~\cite{pielot2017beyond,KongEtAlCHI2018,borkin2015beyond,stokes2022balance}.
Furthermore, considering that bar charts, a commonly used visual representation~\cite{talbot2014four}, tend to elicit biased behavior, we are also motivated to investigate whether textual annotation can mitigate bias in people's behavior when reading bar charts and push them to choose the fair model more often compared to the text-only condition.

\subsection{Hypotheses and Expected Outcomes}
\label{sec:exp3_expected} 

Textual warnings
can mitigate bias in reasoning and decision-making~\cite{pielot2017beyond}. 
However, user interpretation of visualizations can be subconsciously affected by the presence of text, such as slanted titles~\cite{KongEtAlCHI2018}. 
The degree of misalignment between title and visualized data can impact data recall to be more aligned with the title than the data~\cite{kong2019trust}. 
However, textual information does not always overpower visual information.
In a study where participants read captions describing a low-prominence feature in a line chart, they relied on the chart and reported a more visually salient feature as the takeaway~\cite{kim2021towards}.

We hypothesize that explicit textual warnings annotated on bar charts will substantially affect people's perception of model bias as evidenced by changes to their model selections. 
When the text aligns with the bar chart data, we hypothesize that participants will more often select the fair model, even more frequently so than the text condition from Experiment~1.
When the text misaligns with the bar chart data (e.g., the fair model is annotated as the unfair model), prior literature suggests that participants could lean towards either direction: they could rely on the visual information depicted on the bar chart more and behave similarly to the bar condition from Experiment~1, or they could rely on the text more heavily and avoid the model that is annotated as unfair.

\subsection{Participant, Design, and Procedure}

We recruited 212 participants. 
After excluding the participants following the same criteria as before, we ended up with 209 participants. Among them, 101 identified as men ($M_{age}$ = 35.85, $SD_{age}$ = 13.41), 99 identified as women ($M_{age}$ = 39.43, $SD_{age}$ = 13.60), and 9 identified as non-binary or preferred not to disclose ($M_{age}$ = 27.11, $SD_{age}$ = 5.80).

This experiment follows the same procedure and design protocols as Experiment~2 (Section~\ref{sec:Exp2-design}).
Participants read bar charts and select either the fair or the biased model to invest in, across the same four combinations of discrimination values for the biased model as used in Experiment~2 (e.g., men get 65\%, women get 55\%), either for themselves, or on behalf of a gender-unknown client. 
The difference is that participants came across bar charts identical to those used in Experiment~1 (control), or a bar chart with an annotated textual warning above the biased model, or a bar chart with the annotated warning above the fair model --- in a randomized order. Half of the participants invested for themselves, while the other half invested on behalf of a client.
The annotated warning read, ``This robot is unfair to a specific gender''.
We compared participants' perception of fairness when using the default bar chart to make a selection, to that when using bar charts with explicit warnings.
This set-up also allows us to account for the presence of warning overall, and compare the effect of warning alignment to generate insights with regard to how people react when the warning is misaligned with the actual model fairness. 


\subsection{RQ7: Explicit Bias Warning}
\label{sec:explicitbias}

We found a significant main effect of the explicit warning ($F(2, 4752) = 9498.2347, p < 0.001$).  
When there was no explicit warning of bias, $38.0\%$ of the participants chose the fair model. 
Participants became more likely to choose the fair investment model ($54.2\%$) when the biased model was explicitly labeled to be unfair. 
Interestingly, they seem to be significantly impacted by the annotation warning that, when the annotation was paired with the actual fair model, participants were less likely to trust the fair model ($25.4\%$), see Figure~\ref{fig:barexplicit}.

We also replicated findings from Experiment~1 with regard to scenario and gender.
Participants were slightly more likely to choose the fair investment model in the scenario when they were choosing on behalf of a gender-unknown client ($39.4\%$) compared to on behalf of themselves ($39.1\%$), although the effect size is small ($\eta^2_{part} = 3.83 \times 10^{-5}$). 
They were more likely to choose the fair investment model when the difference in return was larger between men and women, despite the model favoring their own gender. $F(3, 4752) = 8269.6615, p < 0.001$. 
The effect of gender also persisted, such that women were more likely to choose the fair investment model. 
Participants who identified as Non-binary choose the fair robot in the conditions when the difference in return was larger (i.e., 85\%/35\%, 35\%/85\%), $49.94\%$ chose fair and $37.04\%$ chose fair in the conditions when the difference was small.

\subsection{Discussion and Summary}

We found support for our hypothesis, such that the warning annotation significantly impacted people's perception of model bias and their selections. 
When the annotation aligned with the bar chart data (e.g., the unfair model is annotated as unfair), participants more often selected the fair model, at an even greater frequency compared to the text condition from Experiment~1.
This suggests that annotating unfair model behaviors can significantly mitigate people's biased interaction with bar charts.
When the annotations misaligned with the bar chart (e.g., the fair model is annotated as unfair), we found participants more heavily relying on the textual information and selected the model \textit{not} annotated as unfair more often.
This corroborates existing work that people rely on textual information more heavily when reasoning with data~\cite{KongEtAlCHI2018, kong2019trust}.
We also replicated the gender and context effects from Experiment~1, such that while women tend to select the fair model more frequently, everyone selected the fair model more often when choosing on behalf of a gender-unknown client.

\section{Design Implications and Future Work}
\label{sec:ImplicationsandFW}

Visualization design has profound consequences on how people perceive fairness and which models they trust. 

\pheading{Design for specific stakeholder-model relationships.} Our work provides preliminary evidence that a stakeholder's relationship to model outcomes, specifically 
investing on one's behalf vs.\ on behalf of a client, affects one's trust decisions (RQ2).  
But with some notable exceptions designed to support MLOps engineers and ML practitioners tackling validation issues~\cite{ZhangEtAlTVCG2023,MunechikaVIS2022}, 
prior work on visualization for fairness assessments~\cite{WexlerTVCG2020, WangEtAlTVCG2021, XieEtAlTVCG2022} has generally not considered this relationship.
We envision this relationship playing a greater role in the design of such systems, explicitly considering which stakeholders will use the tool, and validating the tool accordingly.

\pheading{Embrace the diversity of user perspectives.} How individuals trust models is affected significantly by their demographics (RQ1, RQ5).  
For example, women are more likely to trust fair models than men are. 
Moreover, people follow a broad set of strategies in making trust decisions (RQ6). 
Visualization design must move beyond the monolithic ``user'' and embrace individual differences~\cite{HallTVCG2022, khan2023unbearable}. 

\pheading{Use explicit bias warnings with caution.} Explicitly telling people a model may be biased can overpower the effect of a model's biased history (RQ7). 
While these warnings can enhance communication with the user, erroneous labeling could have a strong detrimental effect. 
The effects of explicit warnings can be more substantial than presentation modality (compare Figures~\ref{fig:barexplicit}~and~\ref{fig:bartextsptablereps}). 
Visualization designers should only use explicit warnings following extensive consultation with stakeholders regarding when and how to deliver the warnings. 

\pheading{Account for designer and user biases.}
The majority of men and women trust and select models biased in their favor (Figure~\ref{fig:exp1_self_client}). 
This result aligns with the neoclassical economists' view that people seek to maximize their profits without considering the effect on others~\cite{GOODLAND198719}. 
In turn, designers must consider and account for how the users' perception of personal advantage will affect their decision-making.
Similarly, the designers' motivations, potential personal gains, and subconscious biases may affect the design and should be explicitly considered.

\pheading{Account for diverse users.}
Gender is not binary, but our study specifically studied the effects of bias against men and women. 
Some participants wondered how the model would perform for someone who does not identify as a man or a woman.
Excluding explicit mentions of potential bias against non-binary individuals led participants who identify as non-binary to employ the indifferent strategy.
This was an unintended flaw in our study design, and our ongoing work is tackling a broader exploration of gender-based biases. 
Designers (and researchers) should consider the implications of presenting biased data for specific groups when the visualizations will be consumed by members of other groups (which has also been done with race~\cite{10.1145/3351095.3372831}). 

\section{Limitations}
\label{sec:Limitations}

\textbf{Participant Limitations:}
All our participants were U.S.-based and received the same compensation, without incurring consequences for their choices.
Our study also focused on bias against men and women, but gender is not binary.
This choice may have adversely affected engagement from certain participants, and future work is needed to understand both the effects on non-binary participants and how to include a more inclusive definition of gender when modeling how bias affects behavior. 
Furthermore, while our study has observed differences in men's and women's behavior, it did not investigate underlying reasoning for those differences. 
Future work should investigate these underlying differences, as well as different participant groups, including additional facets of identify, cultures, and expertise.

\textbf{Study Format Limitations:}
We used a modified instance of a trust game to operationalize trust as the choice between two models, given their historical performance amongst numerous variants of trust games~\cite{BERG1995122, kreps1990corporate, komorita2019social, zheng2002trust}.
Our participants were not given explicit instructions on the interpretation of the history of model returns; they were only told they will see information about how these models have performed in the past, which is consistent with real-world scenarios. Future work could explore how different trust games, presentations, and instruction formats alter trust and interpretations of model behavior.

\textbf{Broader Considerations:}
Our study focused on ML models because that domain routinely exhibits trade-offs between model accuracy and fairness~\cite{Galhotra17fse, Thomas19}; additional work is needed to assess the applicability of our findings to other domains.
Uncertainty can play an important role in visualization and decision-making.
While our studies scratched the surface by showing single average return values via bar charts and textual representations, future work should more explicitly model and study uncertainty and its effect on trust.

\section{Conclusion}
\label{sec:Conclusion}

Our study is the first exploration of how visual design choices, model performance and fairness, and user characteristics affect trust in ML models.  
We identify strategies people use when reasoning about trust in models and find that visual design substantially impacts that trust.
We make concrete recommendations for designing visualization systems that involve ML models. 
Our work is a step towards building empirical visualization knowledge to support ML fairness visualization. 

\section*{Acknowledgments}
The authors wish to thank Alexandra Meliou and Hamza Elhamdadi for their feedback.
This work is supported by the National Science Foundation under grants no.\ IIS-2237585, CCF-2210243, and CCF-1763423, and by Dolby.
Kyle Hall is a Senior Manager at TD in the Global Compliance Department. The content of this ``My Model is Unfair, Do People Even Care? Visual Design Affects Trust and Perceived Bias in Machine Learning'' article reflects Kyle's own opinion and does not represent the views of The Toronto-Dominion Bank.

\balance
\bibliographystyle{abbrv-doi}
\bibliography{references}
\end{document}